\documentclass{article}
\usepackage{amsfonts}
\usepackage{amsmath}
\usepackage{graphicx}

\input{tcilatex}
\begin{document}

\title{The de Broglie Wave as Evidence of\ a Deeper Wave Structure.}
\author{Daniel Shanahan \\
18 Sommersea Drive, Raby Bay, Queensland 4163, Australia}
\maketitle

\begin{abstract}
It is argued that the de Broglie wave is not\ the wave usually supposed, but
the relativistically induced modulation of an underlying carrier wave that
moves with the velocity of the particle. In the rest frame of the particle
this underlying structure has the form of a standing wave. \ 
De Broglie also assumed\ the existence of this antecedent standing wave, 
but it would appear that he failed to
notice its survival as a carrier wave in the Lorentz transformed wave
structure. \ Identified as a modulation, the de Broglie wave acquires a
physically reasonable ontology, evidencing a more natural unity between
matter and radiation than might otherwise be contemplated, and avoiding the
necessity of recovering the particle velocity from a superposition of such
waves. \ Because the Schr\"{o}dinger and other wave equations\ for massive
particles were conceived as equations for the de Broglie wave, this
interpretation of the wave is also relevant to such issues\ in quantum
mechanics as the meaning of the wave function, the nature of wave-particle
duality, and the possibility of well-defined particle trajectories.\medskip

\textbf{Keywords \ }matter wave $\cdot $ wave packet $\cdot $ Schr\"{o}%
dinger equation $\cdot $ wave-particle duality $\cdot $ wave function $\cdot 
$ particle trajectories
\end{abstract}

\section{Introduction}

De Broglie's intuition that solid matter might be wave-like revealed a unity
in Nature that was to play a central role in the formulation of quantum
mechanics (de Broglie \cite{comptes} to \cite{thesis}). \ It was this unity
that allowed all particles, whether massive or massless, to be treated in
like manner as evolving and interfering in accordance with associated wave
characteristics.

However, it will be argued here, essentially from a reconsideration of de
Broglie's thesis of 1924 \cite{thesis}, that the de Broglie or matter wave
is not the independent wave conventionally supposed, but the
relativistically induced modulation of an underlying carrier wave that moves
with the velocity of the particle\footnote{%
The matters to be now considered were dealt with by de Broglie in two brief
sections (I and III) of the opening chapter of his thesis \cite{thesis},
numbering only a dozen or so pages, clearly written and rewarding to close
scrutiny. \ For interesting discussions of the emergence of de Broglie's
ideas, as reflected in his earlier papers, see Medicus \cite{medicus} and
Lochak \cite{lochak}.}. \ In the rest frame of the particle, this underlying
structure has the form of a standing wave\footnote{%
Several earlier such proposals were listed in Shanahan \cite{shanahan};
others include Mellen \cite{mellen} and Horodecki \cite{horodecki}.}. \ 

So regarded, the de Broglie wave is not itself, strictly speaking, the
matter wave of quantum mechanics, but evidences the existence of a deeper
wave structure more deserving of that title. \ If, consistently with special
relativity, this underlying structure were assumed to comprise influences
evolving at the velocity $c$ of light, its existence would imply in turn a
deeper and more natural unity between matter and radiation than could be
contemplated if the only wave associated with a massive particle were its
superluminal de Broglie wave.

This is not to question the significance of de Broglie's wave or the
importance to physics of his famous thesis. \ Einstein remarked that de
Broglie had \textquotedblleft uncovered a corner of the great
veil\textquotedblright\ \cite{veil}, and it was from the thesis that Schr%
\"{o}dinger was led to wave mechanics. \ The Schr\"{o}dinger equation was
itself conceived as a general equation for the de Broglie wave (see Bloch 
\cite{bloch}, and Bacciagaluppi and Valentini \cite{baccia}, esp. Chaps. 2
and 11).

Yet the de Broglie wave and the Schr\"{o}dinger wave function have seemed
curious affairs. \ The de Broglie wave is not only superluminal but
strangely disassociated from the subluminal particle that it is forever
overtaking but never outruns. \ And despite the utility of the wave
function, there has been much debate as to what it actually means. \ 

In his report to the Solvay Conference of 1927, Schr\"{o}dinger observed
that\ this \textquotedblleft $\psi $-function\textquotedblright\ seems to
describe, not a single trajectory, but a \textquotedblleft snapshot ....
with the camera shutter open\textquotedblright\ of all possible classical
configurations (see Bacciagaluppi and Valentini \cite{baccia}, p. 411). \ In
standard quantum mechanics, this superposition is intrinsically
probabilistic, but that view has led to the measurement problem, as well
illustrated by Schr\"{o}dinger's \textit{reductio ad absurdum }of the
unobserved cat that is at once dead and alive (Schr\"{o}dinger \cite{cats}).

However, it has been assumed in quantum mechanics,\ both in its standard
form and in its many reinterpretations, as it was by de Broglie and Schr\"{o}%
dinger, that the de Broglie wave is a wave in its own right. \ How the de
Broglie wave might arise instead from the Lorentz transformation of an
underlying standing wave may be explained shortly as follows. \ The crests
of a standing wave rise and fall as one, each simultaneously with the next
(as shown for a one-dimensional wave in Fig. 1(a)(i) ). \ But according to
special relativity, simultaneity is itself relative. \ In an inertial frame
in which the rest frame of the standing wave is moving, the crests of the
wave are observed to peak, not in unison, but in sequence. \ The standing
wave suffers a dephasing - a failure of simultaneity - in the direction of
travel (as suggested by Fig. 1(a)(ii) ). \ This dephasing, moving through
the underlying wave at superluminal velocity, is the de Broglie wave, not a
wave in its own right, but a modulation, or as it might be termed,
\textquotedblleft a wave of simultaneity\textquotedblright . \ To an
observer for whom the particle wave is moving at the velocity $v$, the
standing wave will have the character of a carrier wave moving at that same
velocity $v$, but subject to a modulation of velocity $c^{2}/v$, where $c$
is the speed of light in vacuum. \ (The significance of Fig. 1(b) will be
explained in Sect. 2.) \ 

\includegraphics[width=7.5cm]{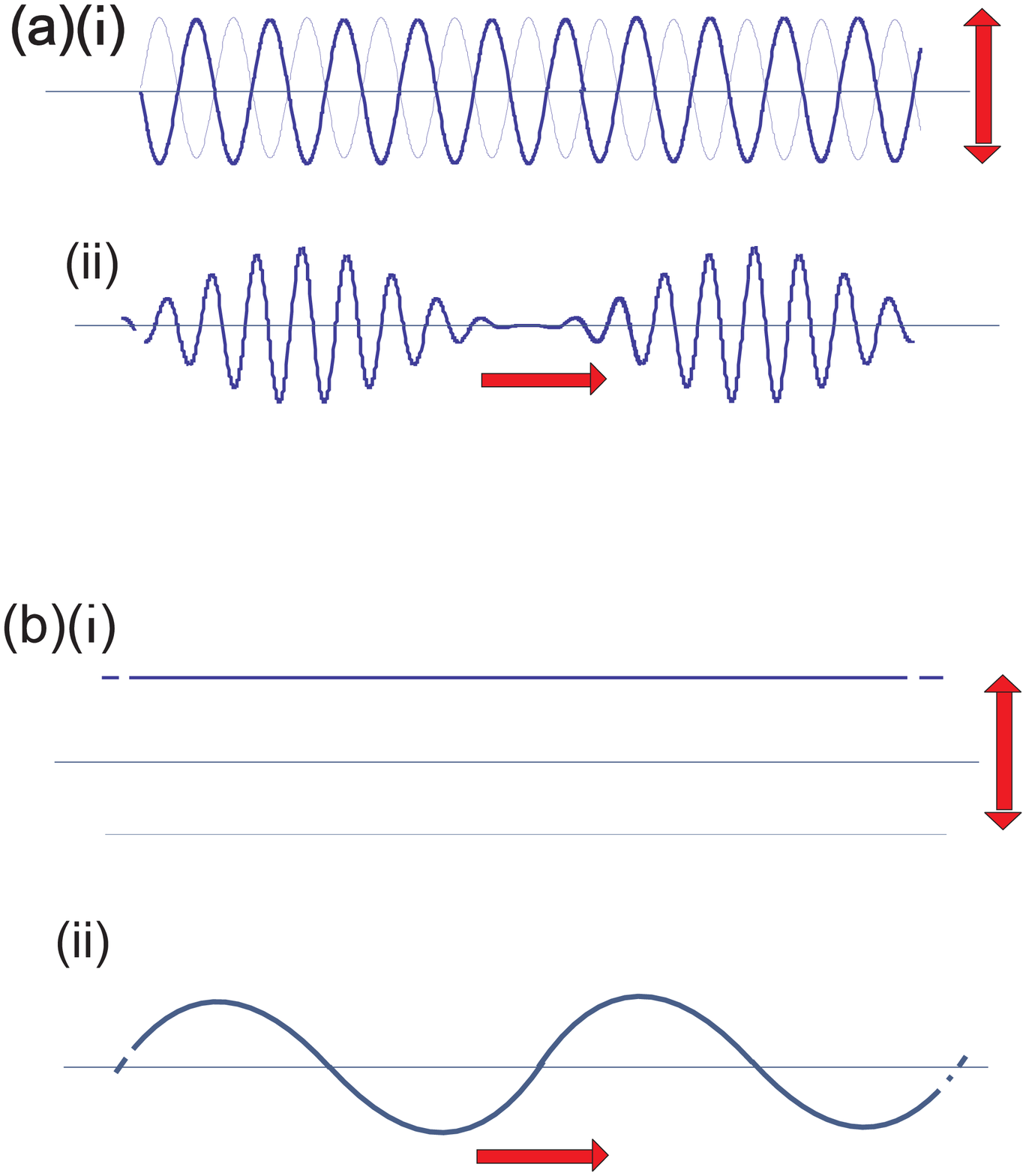}
\begin{quotation}
Fig. 1.\ The Lorentz transformation of two standing waves. \ Standing wave
(a)(i) becomes the modulated wave (a)(ii) comprising a carrier wave of
velocity $v$ subject to a modulation of superluminal velocity $c^{2}/v$. \
The modulation is explained in the text as the de Broglie wave. \ De Broglie
too assumed that the de Broglie wave emerges from a standing wave, but
described a nodeless standing wave of form (b)(i) that has everywhere the
same amplitude. \ When transformed, this wave has the form (b)(ii) in which
the sinusoidal effect evolving at velocity $c^{2}/v$ is again the de Broglie
wave, not an independent wave, but the modulation of a nodeless carrier wave
moving at velocity $v$.
\end{quotation}
\medskip

That this is indeed the correct interpretation of the de Broglie wave is
suggested by the curious manner in which its velocity varies with particle
velocity. \ The velocity of the wave is not only superluminal but increases
as the particle slows and becomes infinite as the particle comes to rest. \
This is not the behaviour of any independent wave, not that at least of any
energy-carrying independent wave, but is precisely the behaviour of the
modulation of an underlying carrier wave.

As we consider de Broglie's thesis, it will become apparent not only that he
too assumed the existence of this antecedent standing wave, but that in two
of his three demonstrations of the de Broglie wave, or as he then called it,
the \textquotedblleft phase wave\textquotedblright \footnote{%
Not the usual sense in which \textquotedblleft phase wave\textquotedblright\
is used, but suggested to de Broglie by the wave's role in the
\textquotedblleft harmony of phases\textquotedblright\ to be discussed in
Sect. 4, below.}, this wave arises exactly as contended here. \ But in view
of the tentative nature of his proposals, and as he explained in the
conclusion to the thesis, he \textquotedblleft left intentionally
vague\textquotedblright\ (\textit{intentionnellement laiss\'{e} assez vagues}%
) the definitions of these wave forms. \ In the absence of a description of
the antecedent standing wave, he seems to have assumed that it simply
becomes the superluminal de Broglie wave when Lorentz transformed. \ It will
be shown that this is not so.

Once the de Broglie wave is identified as a modulation, its velocity becomes
consistent with special relativity, avoiding the necessity of assimilating
the velocity of the particle to the group velocity of a superposition of
such waves. \ With even greater significance for quantum mechanics, the Schr%
\"{o}dinger and other equations for massive particles, including the Pauli,
Klein-Gordon and Dirac equations, which were all contrived from the wave
characteristics of the de Broglie wave (see, for example Dirac \cite{dirac}%
), must then be in some sense equations for a modulation.

This might explain why the wave functions that emerge as solutions of these
equations have seemed incapable of defining a trajectory for the particle in
question. \ The existence of an underlying wave structure, moving at the
velocity of the particle, might also shed light on such issues in quantum
mechanics as the meaning of the wave function and the nature of
wave-particle duality. \ 

De Broglie's interpretation of his wave varied over time, but it was never
the modulation proposed here. \ It was always a wave in its own right, or a
superposition of such waves, as described in the thesis, and it is on the
thesis that we now concentrate. \ It is also from the famous thesis, rather
than\ from de Broglie's later writings, that it is possible to discern where
and how the existence of the underlying carrier wave came to be overlooked.

\section{The \textquotedblleft periodic phenomenon\textquotedblright}

De Broglie saw that if the Planck-Einstein relation, 
\begin{equation}
E=h\nu ,  \label{planck}
\end{equation}%
for the photon were extended to matter and equated with Einstein's
statement, 
\begin{equation}
E=mc^{2},  \label{einstein}
\end{equation}%
of the equivalence of mass and energy, a massive particle could be
associated in its rest frame with a characteristic frequency $\nu _{0}$ or,
when expressed as an angular frequency, 
\begin{equation*}
\omega _{0}=2\pi \nu _{0}=\frac{mc^{2}}{\hbar },
\end{equation*}%
where $m$ and $\hbar $ are respectively the rest mass of the particle and
the reduced Planck constant (de Broglie \cite{thesis}, Chap. 1, Sect I). \ 

De Broglie rejected the possibility that this frequency could be the measure
of some wholly internal oscillation. \ Observing that the energy of an
electron \textquotedblleft spreads throughout all space\textquotedblright ,
he argued that a massive particle must be surrounded in its rest frame by
what he referred to as a \textquotedblleft periodic
phenomenon\textquotedblright . \ 

He clearly regarded this phenomenon as some form of standing wave. \ This is
apparent from his modelling of the phenomenon as an assemblage of springs
oscillating in unison (see Sect. 5 below) and is explicit in his depiction
of the wave in Minkowski spacetime (see Sect. 6 below). \ He expressly
described the phenomenon as a stationary or standing wave in his report to
the Solvay Conference of 1927 (see Bacciagaluppi and Valentini \cite{baccia}%
, p. 341), as also in other works (\cite{structure}, and \cite{intro}, Chap.
3) including his Nobel lecture of 1929 \cite{nobel}. \ Consistently with
that description,\ he suggested elsewhere that in the rest frame of the
particle the de Broglie wave comprises a superposition of incoming and
outgoing waves \cite{frequence}. \ 

But in the thesis itself, de Broglie avoided saying much at all about this
underlying structure. \ As mentioned above, this vagueness was intentional,
the theory being, as he also said in\ concluding the thesis,
\textquotedblleft not entirely precise\textquotedblright\ (\textit{n'est pas
enti\`{e}rement pr\'{e}cis\'{e}}). \ De Broglie did state that the periodic
phenomenon varies sinusoidally in time and that it \textquotedblleft is
distributed throughout an extended region of space\textquotedblright\ (de
Broglie \cite{thesis}, Chap. 1, Sect I). \ But there is no description in
the thesis of the manner in which the wave varies spatially, or any analysis
in mathematical terms of how a standing wave changes under the Lorentz
transformation. \ Nor is consideration given to the constraints imposed by
special relativity on the form that such a wave might take. \ 

By leaving the spatial variation of the phenomenon undefined, de Broglie
allowed the possibility that it is a standing wave with no spatial variation
- a node-free standing wave that has everywhere the same phase and amplitude
(as in Fig. 1(b)(i) ). \ A wave of that form would comprise
counter-propagating waves of infinite wavelength and velocity contrary to
the assumptions of special relativity. \ But it is nonetheless a featureless
standing wave of this kind that de Broglie seems to contemplate in the
thesis, and when Lorentz transformed, this featureless standing wave becomes
(as will be seen in Sects. 5\ and 6\ below) a correspondingly featureless
carrier wave subject to a sinusoidal modulation (as in Fig. 1(b)(ii) ).

In the transformed wave structure, the only \textquotedblleft
wave\textquotedblright\ that is distinguished by a spatial variation is thus
the modulation, which for a particle moving at velocity $v$, evolves through
the carrier wave at the superluminal velocity $c^{2}/v$. \ It became
possible for de Broglie to conclude\ then (or so we surmise) that when
Lorentz transformed the standing wave simply becomes an independent
superluminal wave of velocity $c^{2}/v$. \ 

However, it will be shown in the next section that a standing wave, whatever
its form or frequency, becomes to an observer for whom the rest frame of the
wave is moving at velocity $v$, a carrier wave of that velocity, subject to
a phase modulation having the wave characteristics and superluminal velocity 
$c^{2}/v$ of the de Broglie wave. \ 

\section{The modulation\ }

Consider a standing wave of form, 
\begin{equation}
R(x,y,z)\,e^{i\omega t},  \label{gen}
\end{equation}%
where $R(x,y,z)$ describes the spatial variation of the wave, and $%
e^{i\omega t}$ is its evolution in time.

This wave thus has a well-defined sinusoidal frequency and might be expected
to vary sinusoidally in space as well as time. \ But for now, we will follow
de Broglie in leaving the spatial variation of the wave \textquotedblleft
intentionally vague\textquotedblright . \ 

We want to know the form that this \textquotedblleft periodic
phenomenon\textquotedblright\ would take when observed from a frame of
reference in which the particle is moving with velocity $v$. \ Assuming a
boost in the $x$ direction, and applying the Lorentz transformation, $%
\Lambda _{x}(v)$:%
\begin{gather*}
x^{\prime }=\gamma \left( x-vt\right) , \\
y^{^{\prime }}=y, \\
z^{^{\prime }}=z, \\
t^{\prime }=\gamma \left( t-\frac{vx}{c^{2}}\right) ,
\end{gather*}%
where $\gamma $ is the Lorentz factor,%
\begin{equation*}
(1-\frac{v^{2}}{c^{2}})^{-\frac{1}{2}},
\end{equation*}%
standing wave (\ref{gen}) becomes the moving wave,%
\begin{equation}
R(\gamma \left( x-vt\right) ,\,y,\,z)\,e^{i\omega \gamma \left(
t-vx/c^{2}\right) }.  \label{movgen}
\end{equation}%
which has two wave factors. \ The first,%
\begin{equation}
R(\gamma \left( x-vt\right) ,\,y,\,z),  \label{lorgen}
\end{equation}%
is a carrier wave, which is moving at the velocity $v$ and, as indicated by
the inclusion of the Lorentz factor $\gamma $, has suffered the contraction
of length predicted by special relativity.

The second wave factor,%
\begin{equation}
e^{i\omega \gamma \left( t-vx/c^{2}\right) },  \label{lormod}
\end{equation}%
is a transverse plane wave, which is moving through the carrier wave (\ref%
{lorgen}) at the superluminal velocity $c^{2}/v$. \ Identifying the
frequency $\omega $ with the characteristic frequency $\omega _{0}$ of a
massive particle, wave factor (\ref{lormod}) can be rewritten in terms of
the Einstein frequency,%
\begin{equation}
\omega _{E}=\gamma \omega _{0},  \label{ein}
\end{equation}%
and de Broglie wave number,%
\begin{equation}
\kappa _{dB}=\gamma \omega _{0}\frac{v}{c^{2}},  \label{deb}
\end{equation}%
as,%
\begin{equation}
e^{i(\omega _{E}t-\kappa _{dB}x)},  \label{debr}
\end{equation}%
and is now more clearly identifiable as the de Broglie wave. \ However, it
is not here the independent wave supposed by de Broglie, but as we have
stressed, the modulation of the carrier wave (\ref{lorgen}), defining the
dephasing of that wave (and the failure of simultaneity) in the direction of
travel.

It is the modulated wave,%
\begin{equation}
R(\gamma \left( x-vt\right) ,\,y,\,z)\,e^{i(\omega _{E}t-\kappa _{dB}x)},
\label{compo}
\end{equation}%
rather than the de Broglie wave (\ref{debr}) that displays the full
complement of changes in length, time and simultaneity contemplated by
special relativity. \ It is suggested that it would be anomalous if any
spatially extended phenomenon, wave or otherwise, could be\ Lorentz
transformed into something that did not incorporate all these changes. \
Yet, as will be seen in Sects. 4 to 6, the carrier wave is effectively
suppressed in de Broglie's derivations. \ 

Let us now suppose that instead of a spatially extended wave, we have the
oscillating point or point particle,%
\begin{equation}
\delta \lbrack x_{0},y_{0},z_{0}]\,e^{i\omega _{0}t},  \label{point}
\end{equation}%
where $\delta \lbrack x,y,z]$ is the Dirac delta function and the particle
is located in its rest frame at the point $(x_{0},y_{0},z_{0})$. \ Under a
boost in the $x$ direction,\ oscillating point (\ref{point}) becomes, 
\begin{equation}
\delta \lbrack \gamma (x_{0}-vt),\,y_{0},z_{0}]\,e^{i(\omega _{E}t-\kappa
_{dB}x)},  \label{mpoint}
\end{equation}%
which describes not a wave, but a moving and oscillating point. \ 

Like a child on a pogo stick, this moving point might describe the form of a
wave, but it is not itself a wave. \ It is not possible for a point to
become, by Lorentz transformation, an extended wave. \ Under a Lorentz
transformation, a point remains a point, and a wave, although changed in
form, remains a wave. \ While the second factor in Eqn. (\ref{mpoint}) does
have the functional form (\ref{debr}) of the de Broglie wave, it is not in
this case a wave, and it is not therefore the de Broglie wave.

That this is so may be intuitively obvious, but will be of some importance
in understanding de Broglie's analyses. \ De Broglie was correct in
insisting that if the moving particle is to be associated with a spatially
extended waveform, some such waveform must also exist in the rest frame of
the particle. \ But it will become apparent in the next section (Sect. 4)
that de Broglie disregarded this extended waveform when applying what he
referred to as the theorem of the harmony of phases. \ It will be seen that
by confining\ his consideration of phase to the phase of the particle at the
position of the particle, he derived, not the de Broglie wave, but what was
in effect the variation in phase of a moving and oscillating point.

A difficulty of a different kind will be encountered in Sects. 5 and 6 below
when we consider de Broglie's two further demonstrations of his wave. \ It
is not every wave of the form (\ref{gen}) to which the Lorentz
transformation can be validly applied. \ It was not stipulated when defining
wave (\ref{gen}) that it should comprise underlying influences propagating
at the velocity $c$ of light. \ Yet\ it is that velocity (together with
relative velocity $v$) that determines the velocity $c^{2}/v$ of the
modulation. \ It is the Lorentz transformation itself that imposes the
velocity $c$, and it does so because it is assumed in special relativity
that all underlying influences develop ultimately at that velocity (see, for
example, Ref. \cite{shanahan}, Sect. 6).

Or to put this another way, the de Broglie wave is not merely evidence of
the wave-like nature of matter, but provides confirmation through its
velocity $c^{2}/v$ that the wave-like influences underlying matter and its
interactions evolve ultimately at the free space velocity $c$ of light.

The Lorentz transformation can be appropriately applied to a standing wave
formed from waves of a velocity other than $c$, but only if the velocity of
underlying influences is nonetheless $c$, as is so for instance in the case
of counter-propagating sound or water waves, where changes in underlying
electromagnetic fields propagate at that velocity, or counter propagating
light waves of velocity $c/n$ in a medium of refractive index $n$. \ But the
Lorentz transformation cannot be validly applied to a standing wave of the
form (\ref{gen}) if the spatial variation $R(x,y,z)$ is unphysical, as in
the case of the standing wave that is entirely without spatial variation
referred to in the previous section (Sect. 2). \ Such a wave would be
composed from influences of infinite wavelength and thus infinite velocity,
contrary to special relativity. \ 

But a wave of that kind might nonetheless be simulated, at least in
principle, by an array of identical oscillators. \ Such a simulation will be
encountered in the second of de Broglie's three demonstrations (to be
discussed in Sect. 4), while the unphysical wave simulated will itself be
seen in the third (Sect. 5).

\includegraphics[width=6.5cm]{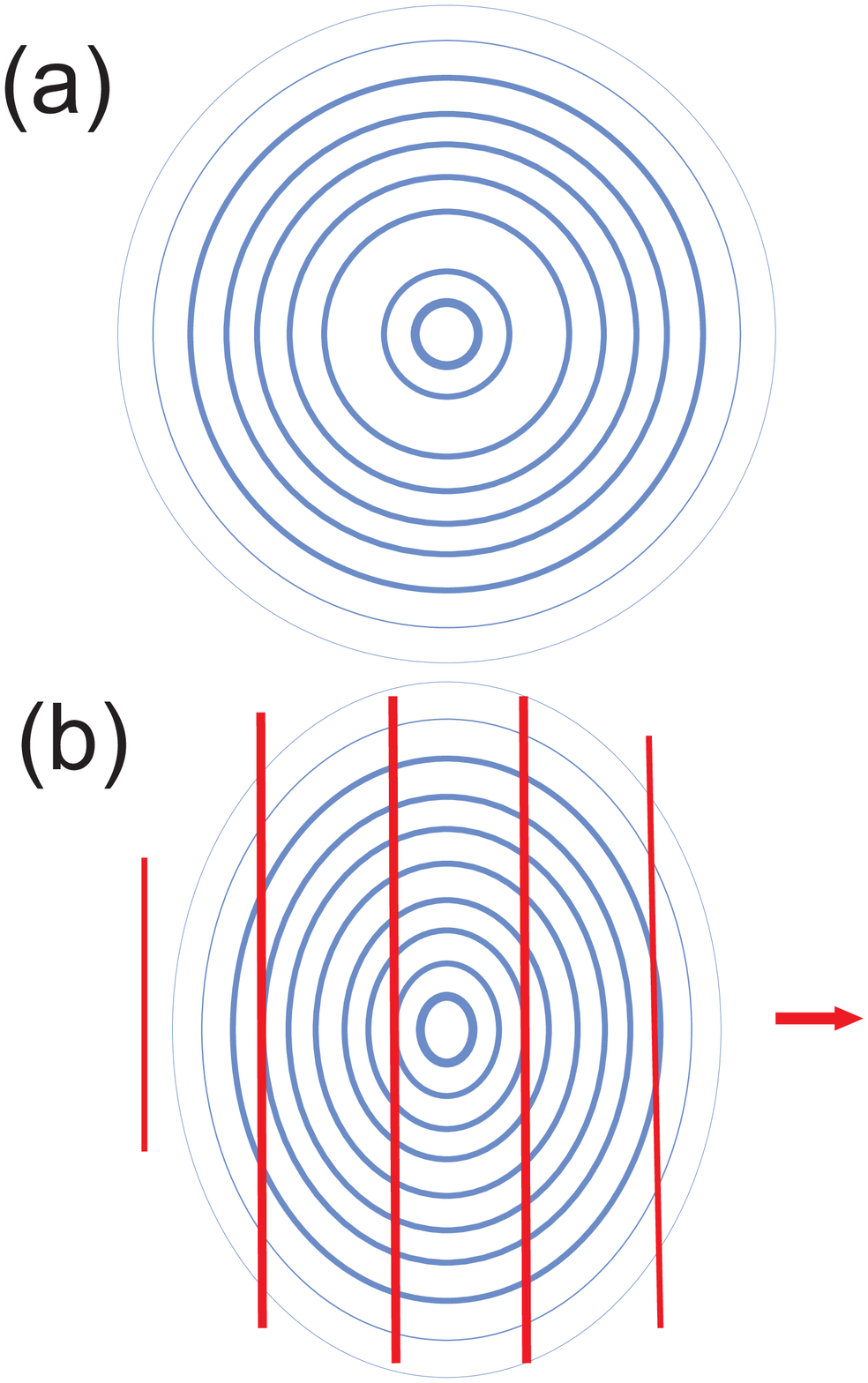}
\begin{quotation}
Fig. 2.\ A model particle wave is represented: (a) as a spherical standing
wave; and (b) as a relativistically contracted carrier wave of velocity $v$,
subject to a modulation (a de Broglie wave) of superluminal velocity $c^{2}/v
$. \ The ellipses in (b) represent the ellipsoidal maxima of the carrier
wave, while the vertical lines are intended to suggest the transverse planar
wave fronts of the de Broglie modulation.
\end{quotation}
\medskip

The relativistic transformation of such an unphysical wave was illustrated
in Fig. 1(b). \ The corresponding transformation of a physically reasonable
standing wave comprising underlying influences of velocity $c$ and varying
sinusoidally in space and time was as shown in Fig. 1(a). \ The Lorentz
transformation of a simple ansatz or model of the wave of Fig. 1(a) in three
dimensions is shown in Fig. 2 \ (see, also Ref.\ \cite{shanahan}).

Finally, it is instructive to apply the reverse Lorentz transformation $%
\Lambda _{x}(-v)$ to the de Broglie wave, 
\begin{equation}
e^{i(\omega _{E}t-\kappa _{dB}x)},
\end{equation}%
whereupon with the aid of Eqns. (\ref{ein}) and (\ref{deb}) it becomes the
oscillating point,%
\begin{equation*}
e^{i\omega _{0}t},
\end{equation*}%
which, as we have stressed, cannot become by Lorentz transformation, a wave.
\ When Lorentz transformed, it describes the track of a moving and
oscillating point. \ 

It has thus been shown\ that the de Broglie wave is not an independent wave,
but the modulation of an underlying carrier wave. \ We now consider in the
next three sections (Sects. 4 to 6) how de Broglie came to conclude
otherwise.

\section{A \textquotedblleft harmony of phases\textquotedblright}

De Broglie's theorem of the harmony of phases (de Broglie \cite{thesis},
Chap. 1, Sect I) is essentially the requirement that relatively moving
observers should agree on the phase that a wave has at each point of space
and time. \ Phase is a scalar invariant and must have the same value in all
inertial frames.

De Broglie concluded on the basis of this theorem that a moving particle is
accompanied by a superluminal wave of velocity $c^{2}/v$. \ But in reaching
that conclusion he confined his harmonizing of phase to the phase of the
particle at the position of the particle. \ Thus what he was considering was
not a wave but the sinusoidal trace of an oscillating point as defined by
Eqn. (\ref{mpoint}).

De Broglie's argument proceeded as follows: \ From the standpoint of a
\textquotedblleft fixed observer\textquotedblright\ (for whom the particle
is moving at velocity $v$) the particle has in its rest frame the reduced
frequency, 
\begin{equation}
\omega _{red}=\frac{\omega _{0}}{\gamma },  \label{wred}
\end{equation}%
yet in the frame of that same observer, the moving particle has an increased
energy and correspondingly increased frequency,%
\begin{equation}
\omega _{inc}=\omega _{0}\gamma .  \label{winc}
\end{equation}

But a wave can have only one phase at any point of space and time, and all
observers must agree on that phase. \ As de Broglie put it\footnote{%
In the quoted passage, we have simplified de Broglie's expressions for the
frequencies and avoided here as elsewhere in this paper the practice of
describing the velocity of the de Broglie wave as $c/\beta $.},

\begin{quotation}
The periodic phenomenon connected to a moving body whose frequency is for
the fixed observer equal to [$\omega _{0}/\gamma $] appears to him to be
constantly in phase with a wave of frequency [$\omega _{0}\gamma $] emitted
in the same direction as the moving body, and with the velocity $V=$[$%
c^{2}/v $].
\end{quotation}

So far so good. \ De Broglie refers in this passage to the periodic
phenomenon and appears to be contemplating the transformation of the entire
spatially extended phenomenon. \ But when he then proceeds to derive the
velocity $V=c^{2}/v$ of the resulting \textquotedblleft
wave\textquotedblright , he confines his consideration to a single point
within the phenomenon, that is to say, the position of the particle, which
is stationary in one frame and moving in the other, but is nonetheless a
single point developing along a single world line.

De Broglie states (correctly) that as the particle travels the distance $x$
in the time $t$ (in the frame of the fixed observer), it follows from Eqn. (%
\ref{wred}) that it is considered to experience in its own frame the change
of phase,%
\begin{equation}
\omega _{red}\,t=\frac{\omega _{0}}{\gamma }\frac{x}{v},  \label{w1}
\end{equation}%
whereas in the fixed observer's frame, it follows from Eqn. (\ref{winc})
that the change of phase observed by that observer is,%
\begin{equation}
\omega _{inc}\,(t-\frac{x}{V})=\omega _{0}\gamma \,(\frac{x}{v}-\frac{x}{V}).
\label{w2}
\end{equation}%
These changes of phase must be equal. \ Thus, equating (\ref{w1}) and (\ref%
{w2}),%
\begin{equation*}
\frac{\omega _{0}}{\gamma }\frac{x}{v}=\omega _{0}\gamma (\frac{x}{v}-\frac{x%
}{V}),
\end{equation*}%
from which,%
\begin{equation*}
V=\frac{c^{2}}{v}.
\end{equation*}

De Broglie's analysis thus delivers the velocity of the de Broglie wave. \
But the change described by Eqn. (\ref{w2}) is simply the change in phase
that occurs in a point particle as it moves the distance $x$ in the time $t$%
. \ In terms of the Einstein frequency $\omega _{E}$ and de Broglie wave
number $\kappa _{dB}$, the evolution of phase $\Delta \phi $ described by
Eqn. (\ref{w2}) can be rewritten using Eqns. (\ref{ein}) and (\ref{deb}) as,%
\begin{equation}
\Delta \phi =\omega _{E}\,t-\kappa _{dB}\,x,  \label{debphase}
\end{equation}%
\ from which it can be seen (compare, for instance, Eqn. (\ref{debr}) ) that
the oscillating point, moving at the subluminal velocity $v$, is maintaining
consistency of phase with a superluminal wave having the characteristics of
the de Broglie wave, which\ explains why de Broglie's analysis was able to
produce the velocity of the de Broglie wave.

But by seizing upon the single point, and ignoring the disposition of phase
across the extended wave, de\ Broglie suppressed the carrier wave of
velocity $v$, which as shown in Sect. 3, must result from the Lorentz
transformation of a standing wave. \ If a wave with the characteristics of
the de Broglie wave were the only wave associated with the moving particle,
harmony of phase could be guaranteed only at the position of the particle. \
It is the full modulated wave that harmonizes phase at all points for all
observers.

De Broglie's two other demonstrations did involve the transformation of an
extended wave. \ But from his theorem of the harmony of phases, he had
already concluded that the de Broglie wave is a wave in its own right.

\section{The mechanical model}

Faced with the potentially embarrassing superluminality of his wave, de
Broglie invoked a simple toy model to illustrate how a velocity greater than 
$c$ might yet be consistent with special relativity provided the actual
velocity of energy transport were less than $c$ (de Broglie \cite{thesis},
Chap. 1, Sect I). \ \ 

\includegraphics[width=9.5cm]{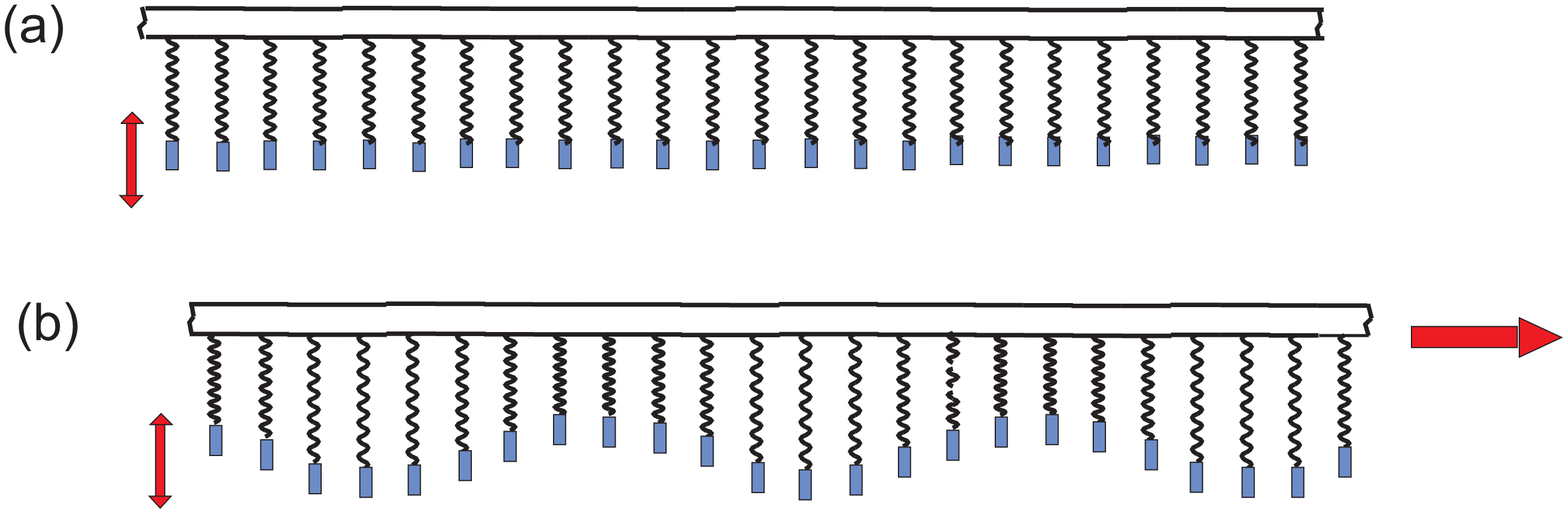}

\begin{quotation}
Fig. 3. De Broglie's mechanical model shown stationary in (a), and moving at
the relativistic velocity $v$ in (b). \ The sinusoidal effect of
superluminal velocity $c^{2}/v$ is a consequence of the failure of
simultaneity in the direction of travel. \ This sinusoidal "wave" is the de
Broglie wave, not an independent wave as assumed by de Broglie, but a
dephasing (a modulation) of the standing wave modelled by the assemblage of
springs.
\end{quotation}

Thus de Broglie was not seeking to derive his wave from this model, merely
to justify its superluminal velocity. \ Even so, the de Broglie wave emerges
from the model, not as an independent wave, but as the modulation of an
underlying carrier wave.

As described by de Broglie, and as we have imagined this model in Fig. 3, it
comprises a horizontal disk of large diameter, from which are suspended many
identical spring weights oscillating in phase and at the same amplitude, but
with the number of springs per unit area diminishing with distance from the
centre of the disc in \textquotedblleft very rough analogy\textquotedblright
, as de Broglie explained, to the distribution of energy around a particle.
\ (But no attempt has been made to depict this diminishing intensity in Fig.
3).

An observer in the inertial frame of the disk observes these weights to be
oscillating in unison (as in Fig.\ 3(a), and see also again, Fig. 1(b)(i) ).
\ But a second observer, for whom the disk is moving at velocity $v$ (Figs.
3(b) and 1(b)(ii) ), observes (along with other relativistic effects) what
de Broglie described as the \textquotedblleft dephasing of the movements of
the various weights\textquotedblright , that is to say, the failure of
simultaneity in the direction of travel. \ To the first observer, the
weights define a horizontal plane moving up and down. \ But to the observer
for whom the disc is moving, this surface is not planar but sinusoidal, with
the crests of this sinusoidal surface moving in the same direction as the
disk, but at the superluminal velocity $c^{2}/v$ of the de Broglie wave.

De Broglie did not provide a separate derivation of the velocity $c^{2}/v$
of this sinusoidal effect. \ He merely said of the moving sinusoidal surface
that:

\begin{quotation}
It corresponds .... to our phase wave. \ According to the general theorem,
the surface has a speed [$c^{2}/v$] parallel to that of the disk ... With
this example we see clearly (and this is our excuse for such protracted
insistence on it) how the phase wave corresponds to the transport of the
phase and not at all to that of the energy.
\end{quotation}

In other words, the sinusoidal surface defined by the moving springs is an
instance of his\ phase wave (the de Broglie wave), while the
\textquotedblleft general theorem\textquotedblright\ de Broglie relies upon
is his theorem of the harmony of phases. \ 

De Broglie seems not to have noticed that the standing wave (the array of
oscillating weights) has not simply become the de Broglie wave. \ It has
become a structure moving at velocity $v$ (a moving array of oscillating
weights) subject to a modulation moving at velocity $c^{2}/v$. \ In treating
the modulation as an independent wave, de Broglie has ignored the structure
that it modulates.

In its rest frame, this toy model is not strictly speaking of course a
standing wave. \ It is merely the \textit{simulation} of a standing wave; in
fact, the simulation of a physically unreasonable nodeless standing wave
that has everywhere the same phase and amplitude. \ While an assemblage of
synchronized oscillators would not itself be physically unreasonable, the
wave thus modelled would comprise underlying influences of infinite velocity
contrary to special velocity. \ But even ignoring that difficulty, it was
shown in Sect. 3 that the application of the Lorentz transformation to a
standing wave of any form results formally, not in an independent wave of
velocity $c^{2}/v$, but in a carrier wave of velocity $v$ subject to a
modulation of velocity $c^{2}/v$. \ 

De Broglie's model becomes under a Lorentz transformation, not the
simulation of an independent de Broglie wave, but the simulation of a
modulated wave of velocity $v$ in which the de Broglie wave is the
modulation.

If de Broglie had supposed an assemblage of springs that varies sinusoidally
in space as well as time (and thus of the general form depicted in Fig. 1(a)
), and in which underlying influences move at velocity $c$, he would have
had in that model, a de Broglie wave that is physically reasonable and
consistent with special relativity. \ But when he sought to recover the
classical velocity of the particle from his superluminal wave, the analogy
he drew was with the group velocity of a wave packet formed from the
superposition of de Broglie waves of nearly equal frequency (see de Broglie 
\cite{thesis}, Chap 1, Sect II). \ 

It was from that very different physical\ effect that the difficult concept
of a particle wave packet was carried into quantum mechanics.

\section{In Minkowski spacetime}

By representing a standing wave in a spacetime diagram, de Broglie was able
to derive the velocity of the de Broglie wave, while demonstrating in an
intuitive manner the dephasing and consequent failure of simultaneity
defined by this wave.

\includegraphics[width=10.0cm]{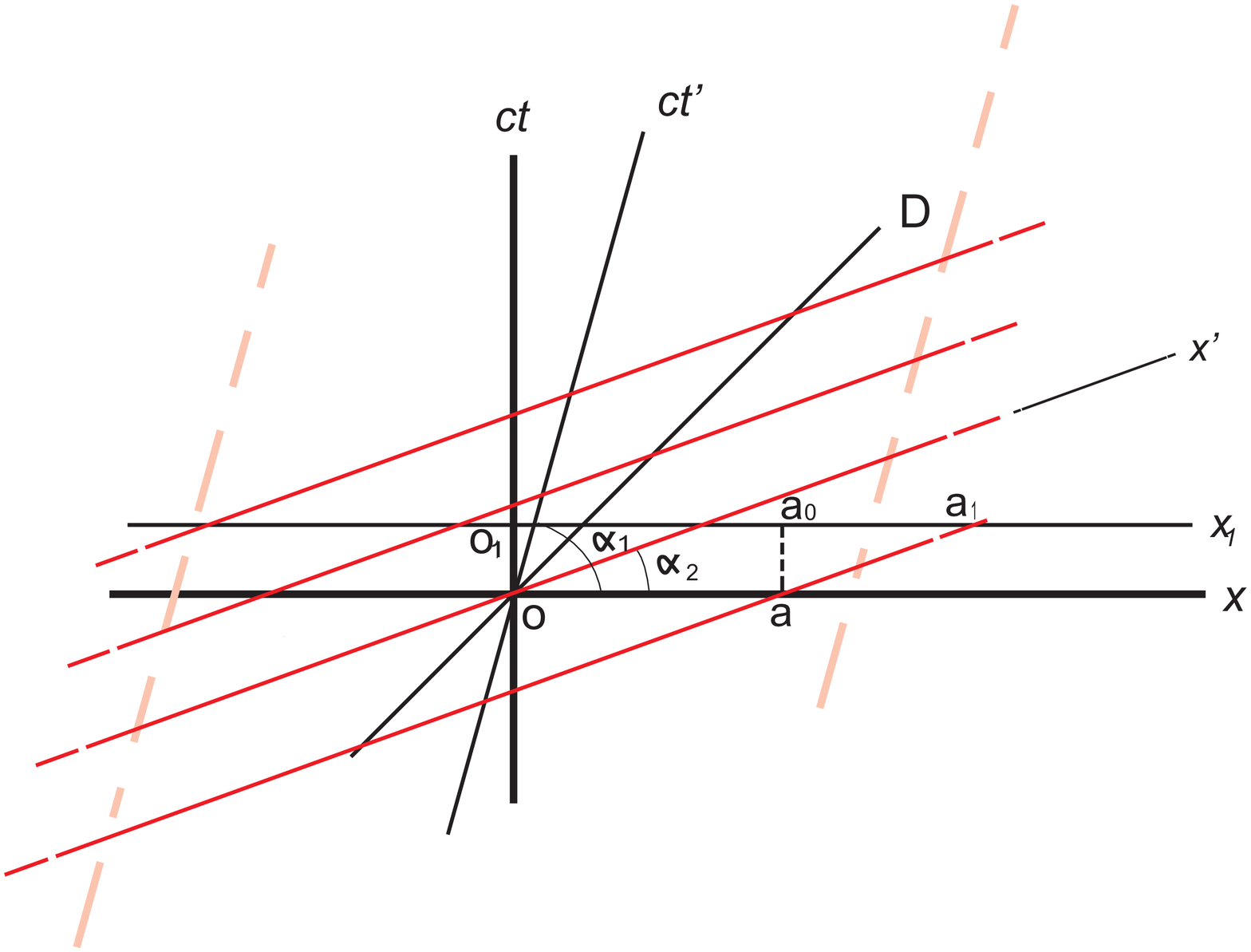}
\begin{quotation}
Fig. 4.\ The world line of a particle follows the $ct^{^{\prime }}$ axis. \
Surrounding the particle in its rest frame is a standing wave represented by
four parallel \textquotedblleft equiphase planes\textquotedblright . \ A
world tube enclosing the world line of the particle has been defined by
dashed lines parallel to the $ct^{^{\prime }}$.axis. \ The equiphase planes
are inclined at $\alpha _{2}$=$\arctan v/c$ to the $x$ axis, and define in
the unprimed frame, a dephasing of velocity $c^{2}/v$. \ This dephasing is
the de Broglie wave, considered by de Broglie an independent wave, but
revealed by the drawing to be a modulation of the transformed standing wave.
\ (Adapted from de Broglie \cite{thesis}, Chap. 1, Sect III, Fig. 1).
\end{quotation}
\medskip

Referring to Fig. 4, the unprimed ($x$, $ct$) coordinates are those of the
fixed observer, while the primed ($x^{^{\prime }}$, $ct^{^{\prime }}$)
coordinates define the frame of the moving particle. \ The $ct^{^{\prime }}$
and $x^{^{\prime }}$ axes are inclined at $\alpha _{1}=\arctan c/v$ \ and $%
\alpha _{2}=\arctan v/c$, respectively, to the $x$ axis, while the line $OD$
defines one edge of the light cone. \ The particle itself is moving to the
right at the velocity $v$, and its world line thus follows the primed $%
ct^{^{\prime }}$\ axis. \ 

The standing wave is represented in the diagram by what de Broglie referred
to as \textquotedblleft equiphase spaces\textquotedblright\ (\textit{espaces 
\'{e}quiphases}), these being the four equally spaced lines drawn parallel
to the $x^{^{\prime }}$ axis. \ De Broglie also referred to these as
\textquotedblleft planes\textquotedblright\ (\textit{plans \'{e}quiphases}),
presumably hyperplanes of three dimensions in the four dimensions of
spacetime. \ Each such plane defines a sub-space in which the wave has
reached a particular phase. \ In each inertial frame, these planes thus
repeat after a time equal to the period of oscillation in that frame.

The four equiphase planes are parallel to the $x^{^{\prime }}$ axis, leaving
no doubt that it was assumed by de Broglie that in the rest frame of the
particle the periodic phenomenon comprises some form of standing wave. \
What we want to know is what these planes mean in the unprimed frame of the
fixed observer.

However, as they appear in the thesis (see de Broglie \cite{thesis}, Chap.
1, Sect. 3, Fig. 1), these planes are depicted in a way that could be a
source of confusion. \ The lines representing the planes are not centred, as
would be natural, on the world line of the particle. \ Indeed they are so
positioned that they could suggest a wave propagating to the left of the
diagram. \ The four planes have thus been repositioned in Fig. 4 so that
they retain the same inclination to the $x$ axis as in de Broglie's drawing
but are centred on the world line of the particle. \

To show how these planes would be represented from the standpoint of an
observer who is in the frame of the particle itself,\ the roles of primed
and unprimed frames have been interchanged in a further diagram (Fig. 5). \
In this, it is the formerly fixed observer, rather than the particle, that
is moving to the right. \ Here again the planes are centred on the world
line of the particle.

However, we concentrate now on Fig. 4. \ In the unprimed frame, the
equiphase planes are inclined to the $x$ axis and thus display the asymmetry
in phase and failure of simultaneity in the direction of travel predicted by
special relativity. \ From this asymmetry, de Broglie calculated the
velocity of the de\ Broglie wave. \ He first explains that as the line $%
O_{1}x_{1}$ represents the frame of the fixed observer at $t=1$, the
distance $aa_{0}$ is exactly $c$. \ He then says that at $t=1$,

\begin{quotation}
The phase which at time $t=0$ was at $a$ is now at $a_{1}$. \ For the fixed
observer it is then displaced in space by an amount $a_{0}a_{1}$ in the $x$
direction, during one unit of time. \ One can then say that the speed is:%
\begin{equation*}
V=a_{0}a_{1}=aa_{0}\cot \alpha _{2}=\frac{c^{2}}{v}.
\end{equation*}%
\smallskip
\end{quotation}

This is again the velocity of the de Broglie wave, but when de Broglie
refers in this passage to the speed at which phase has become
\textquotedblleft displaced in space\ .... in the $x$ direction%
\textquotedblright , he is describing the velocity of a dephasing, that is
to say, of a modulation. \

\includegraphics[width=10.0cm]{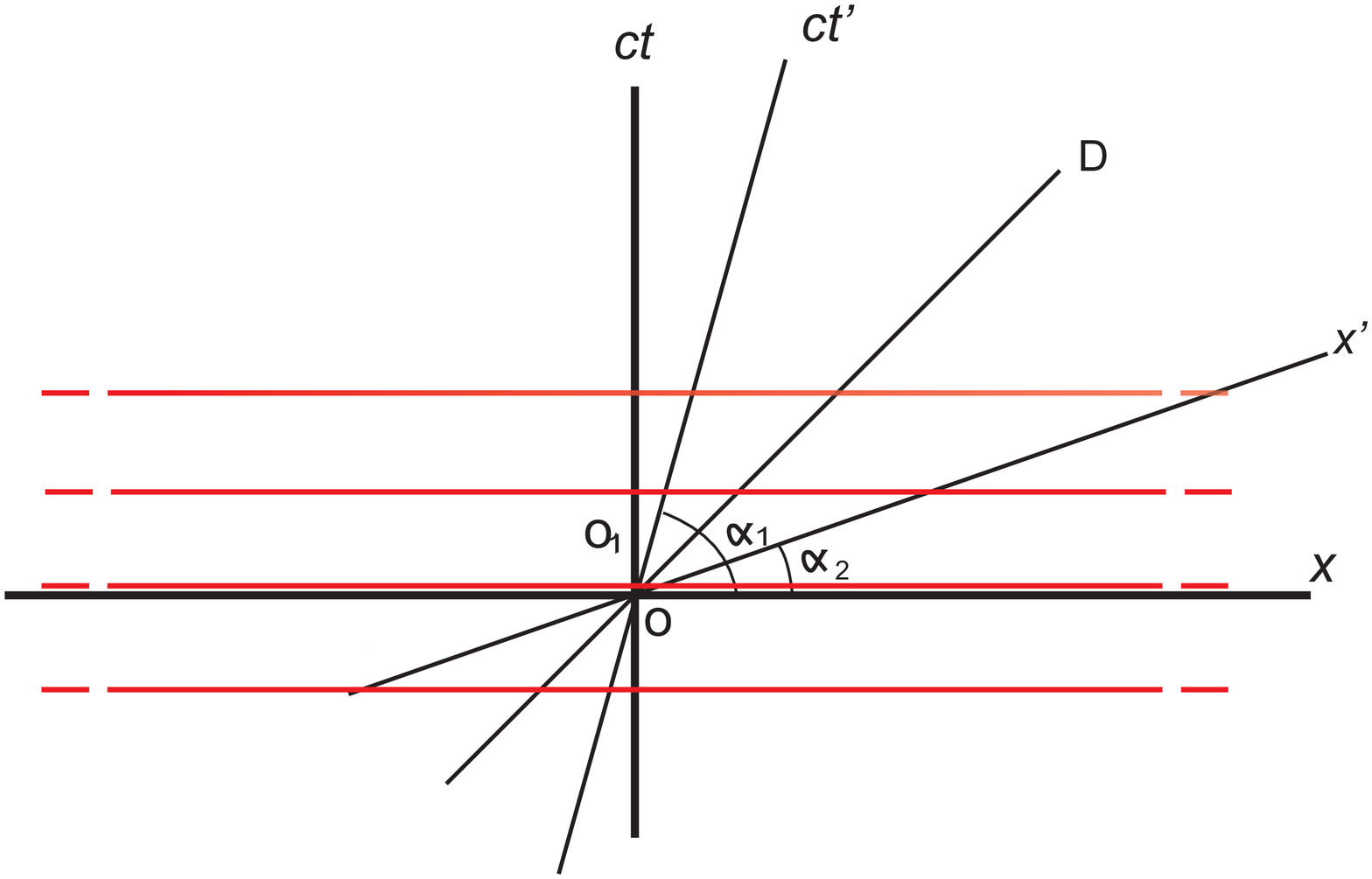}
\begin{quotation}
Fig. 5.\ As Fig. 4, but with the roles of primed and unprimed frames
interchanged. \ The formerly fixed observer is moving to the right with
velocity $v$, while the unprimed frame is now the inertial frame of the
particle. \ The four equiphase planes representing the standing wave\ are
thus parallel to the $x$ axis.
\end{quotation}
\medskip

To see that the standing wave\ has not simply become transformed into an
independent de Broglie wave as supposed by de Broglie, it is only necessary
to notice that the standing wave and the de Broglie wave do not have the
identical representation in the diagram. \ In a spacetime diagram, such as
those included here, the Lorentz transformation is a \textit{passive}
transformation under which an event or world line retains its location
within the diagram\footnote{%
To show the Lorentz transformation as \textit{active}, the diagram would
include two depictions of the equiphase planes, that shown in Fig. 4 and
that shown in Fig. 5.}. \ Instead of the event or world line changing
position, its coordinates differ according to the frame to which it is
referred. \ Thus, if the de Broglie wave and the transformed standing wave
were the same wave they would share the same world line in the drawings. \ 

But that is not so. \ Consider the standing wave. \ In its rest frame, a
standing wave is, by definition, stationary. \ The world line of any point
in that wave thus follows the time axis of its rest frame, which in this
case is the $ct^{^{\prime }}$ axis of the primed frame of Fig. 4. \ 

Or we could consider instead some extended region within the wave, let us
say a spherical region enclosing the particle. \ Instead of a world line,
the passage through spacetime of such a region defines a \textquotedblleft
world tube\textquotedblright\ (see, for example, Misner, Thorne and Wheeler 
\cite{misner}, p. 473), which in this case, and as shown in Fig. 4, would
enclose the world line of the particle\ which is following the primed $%
ct^{^{\prime }}$ axis.

On the other hand, the superluminal velocity $c^{2}/v$ of the de Broglie
wave implies a world line (or tube, but it will suffice now to speak of a
line) lying beyond the light cone and parallel to the $x^{^{\prime }}$ axis.
\ Thus the standing wave does not simply become the de Broglie wave, as was
assumed by de Broglie.

One further feature of Fig. 4 should be noticed. \ The fact that the world
line of the de Broglie wave is parallel to the $x^{^{\prime }}$ axis (or, in
Fig. 5, the $x$ axis) means that in the rest frame of the particle, this
wave is of infinite velocity. \ That velocity would be anomalous in an
independent wave, but acquires a natural explanation once the de Broglie
wave is understood as a modulation. \ At rest, the crests of the underlying
wave are no longer peaking in sequence, but in unison, simultaneity has been
restored, alignment of phase has become instantaneous, and the velocity of
the modulation describing the progress of that alignment has thus become
infinite. \ 

In effect, modulation and carrier have merged in the standing wave, and the
de Broglie wave has disappeared.

\section{Discussion}

There is in de Broglie's thesis an evocative and much-cited passage that
captures the essence of his notion of wave-particle duality:

\begin{quotation}
The particle glides on its wave, so that the internal vibration of the
particle remains in phase with the vibration of the wave at the point where
it finds itself (de Broglie \cite{thesis}, Chap. 1, Sect. 1).
\end{quotation}

And this in a sense is true. \ As discussed in Sect. 4 above, the particle,
considered as a subluminal moving point, changes phase (see Eqn. (\ref%
{debphase}) ) as if it were maintaining consistency of phase with a
superluminal wave having the characteristics of an independent de Broglie
wave. \ But in referring to an \textquotedblleft internal
vibration\textquotedblright\ de Broglie seems to deny the spatially extended
standing wave from which, as we saw in Sect 2, he commenced his analysis.

This and other references in de Broglie's writings to the oscillation of
frequency $\omega _{0}$ being \textquotedblleft intrinsic\textquotedblright\
or \textquotedblleft internal\textquotedblright\ or even \textquotedblleft
fictitious\textquotedblright\ (see, for instance, de Broglie \cite{double})
may be explained by de Broglie's insistence that wave and particle are
ontologically distinct entities. \ But it has been shown here that if the
characteristic frequency $\omega _{0}$ of the particle 
is taken to be that of a standing wave, as de
Broglie himself proposed, the de Broglie wave arises as an immediate
consequence of the failure of simultaneity described by the Lorentz
transformation. \ 

In none of de Broglie's demonstrations in the thesis is the de Broglie wave
the independent superluminal wave that de Broglie contemplated. \ In the
first, harmonizing of phases occurs only for an oscillating point within
each wave and not for the wave as a whole. \ The result of Lorentz
transforming that oscillating point is not a spatially extended wave, but
the path described by a moving oscillating point. \ In the second (the
mechanical model) and the third (the Minkowski diagram), the de Broglie wave
emerges from the antecedent standing wave, not as a wave in its own right,
but as the modulation of an underlying standing wave.

The reinterpretation of the de Broglie wave as a modulation would not affect
the formal core of quantum mechanics. \ The de Broglie wave or wave function
would still evolve in accordance with the wave characteristics determined by
the relevant wave equation. \ But much that has seemed anomalous would be
explained. \ The superluminal velocity of the wave would not then be that of
energy (or information) transport, consistency with special relativity would
be achieved\ and there would be no need to equate the velocity of a massive
particle with the group velocity of a superposition of such waves. \ 

Nor would it be at all mysterious that the superluminal wave does not outrun
the subluminal particle or fly off at a tangent from a particle orbit. \ A
modulation must remain forever coextensive with the wave it modulates. \ As
the particle changed direction, there would simply be a corresponding change
in the direction of dephasing and a rearrangement of phase throughout the
modulated wave. \ This rearrangement of phase would unfold not at the
superluminal velocity of the de Broglie, but at the velocity of the
constituent influences of the underlying wave, which for consistency with
special relativity we assume to be the velocity $c$ of light. \ 

But it is the suggestion of a deeper wave structure underlying the de
Broglie wave that could have the larger significance for quantum theory. \
In the\ mystery of wave-particle duality, the role of wave has been played
solely by the de Broglie wave. \ If the de Broglie wave is the modulation of
a underlying standing wave structure, it becomes possible to speculate that
this underlying wave, moving at the velocity of the particle and following a
well-defined trajectory, might explain both the wave and the particulate
properties of the particle. \ \ 

And composed of influences evolving at the speed of light, the existence of
this underlying structure would imply a deeper and more natural unity in
Nature than could be guessed at if the only wave associated with a massive
particle were of superluminal velocity and unknown ontology. \ \ 


\medskip
\medskip

\begin{thebibliography}{99}
\bibitem{comptes} L. de Broglie, Ondes et quanta, Comptes Rendus, \textbf{177%
}, 507 (1923)

\bibitem{phil} L. de Broglie, A Tentative Theory of Light Quanta, Phil. Mag. 
\textbf{47}, 446 (1924)

\bibitem{thesis} L. de Broglie, Doctoral thesis. Recherches sur la th\'{e}%
orie des quanta. Ann. de Phys. (10) \textbf{3}, 22 (1925). \ For English
translations, see J. W. Haslett, \ Phase waves of Louis de Broglie, Am. J.
Phys. 40, 1315 (1972) (Chap. 1 only); and A. F. Kracklauer, On the Theory of
Quanta,\ http://aflb.ensmp.fr

\bibitem{medicus} H. A. Medicus, Fifty years of matter waves, Physics Today,
Feb. 1974

\bibitem{lochak} G. Lochak, The Evolution of the Ideas of Louis de Broglie
on the Interpretation of Quantum Mechanics, Found. Phys. \textbf{12}, 932
(1982)

\bibitem{shanahan} D. Shanahan, A Case for Lorentzian Relativity, Found.
Phys. \textbf{44}, 349 (2014)

\bibitem{mellen} W. R. Mellen, Moving Standing Wave and de Broglie Type
Wavelength. Am. J. Phys. \textbf{41}, 290 (1973)

\bibitem{horodecki} R. Horodecki, Information Concept of the Aether and its
application in the Relativistic Wave Mechanics, in L. Kostro, A. Poslewnik,
J. Pykacz, M. \.{Z}ukowski (Eds.), Problems in Quantum Physics, Gdansk '87,
World Scientific, Singapore, (1987)

\bibitem{veil} Letter dated 16 Dec. 1924 from A. Einstein to P. Langevin,
quoted in W. J. Moore, Schr\"{o}dinger: Life and Thought, Cambridge
University Press, Cambridge, U.K. (1989), p. 187

\bibitem{bloch} F. Bloch, Heisenberg and the early days of quantum
mechanics. Physics Today, Dec. 1976, adapted from a talk given on 26 April
1976 at the Washington D. C. meeting of The American Physical Society

\bibitem{baccia} G. Bacciagaluppi, A. Valentini, Quantum theory at the
crossroads: Reconsidering the 1927 Solvay Conference. Cambridge University
Press. Cambridge (2009)

\bibitem{cats} E. Schr\"{o}dinger, Die gegenw\"{a}rtige Situation in der
Quantenmechanik, Naturwissen. \textbf{23}, 807 (1935), translated in J. A.
Wheeler, W. H. Zurek, (eds.): \textit{Quantum Theory and Measurement }%
(Princeton University Press, New Jersey, 1983)

\bibitem{dirac} P. A. M. Dirac, The Quantum Theory of the Electron. Proc.
Roy. Soc. A \textbf{117}, 610 (1928)

\bibitem{structure} L. de Broglie, La m\'{e}canique ondulatoire et la
structure atomique de la mat\`{e}rie et du rayonnement. \ Journal de
Physique, \textbf{8}, 225 (1927)

\bibitem{intro} L. de Broglie, An Introduction to the Study of Wave
Mechanics, E. P. Dutton \& Co., New York (1930)

\bibitem{nobel} L. de Broglie, The Wave Nature of the Electron. Nobel
lecture, 12 Dec. 1929. \ In Nobel Lectures, Physics\ 1922-1941, Elsevier,
Amsterdam, 1965

\bibitem{frequence} L. de Broglie, Sur la fr\'{e}quence propre de l'\'{e}%
lectron, C. R. Acad. Sci. \textbf{180}, 498 (1925)

\bibitem{misner} C. W. Misner, K. S. Thorne, J. A. Wheeler,\ Gravitation,
Freeman, New York (1973)

\bibitem{double} L. de Broglie, Interpretation of quantum mechanics by the
double solution theory, Ann. Fond. Louis de Broglie, \textbf{12}, 1 (1987).
\ 
\end{thebibliography}
\end{document}